\documentclass[%
 reprint,
 amsmath,amssymb,
 aps,
showkeys
]{revtex4-2}

\usepackage{graphicx}% Include figure files
\usepackage{dcolumn}% Align table columns on decimal point
\usepackage{bm}% bold math
\usepackage{multirow}
\usepackage{siunitx}
\usepackage{soul}
\usepackage[export]{adjustbox}

\begin{document}

\preprint{APS/123-QED}

\title{Holographic Thermal Mapping Using Acoustic Lenses}% Force line breaks with \\

\author{Ceren Cengiz}
%\email{cerencengiz@vt.edu}
\author{Shima Shahab}
\email{sshahab@vt.edu}
\affiliation{Department of Mechanical Engineering, Virginia Tech, Blacksburg, VA, $24061$, USA}

\date{\today}% It is always \today, today,
             %  but any date may be explicitly specified

\begin{abstract}

Acoustic holographic lenses (AHLs) show great potential for sound manipulation. These lenses store the phase and amplitude profile of the desired wavefront when illuminated by a single acoustic source to reconstruct focused ultrasound (FUS) pressure fields, induce localized heating, and achieve temporal and spatial thermal effects in acousto-thermal materials like polymers. The ultrasonic energy is transmitted and focused by AHL from a transducer into a particular focal volume. It is then converted to heat by internal friction in the polymer chains, causing the temperature of the polymer to rise at the focus locations while having little to no effect elsewhere. This one-of-a-kind capability is made possible by the development of AHLs to make use of the translation of attenuated pressure fields into programmable heat patterns. However, the acousto-thermal dynamics of AHLs are largely unexplored. We use a machine learning-assisted single inverse problem approach for rapid and efficient AHLs’ designs. The process involves the conversion of thermal information into a holographic representation through the utilization of two latent functions; pressure phase and amplitude. Experimental verification is performed for pressure and thermal measurements. The volumetric acousto-thermal analysis of experimental samples is performed to offer a knowledge of the obtained pattern dynamics, as well as the applicability of holographic FUS thermal mapping for precise temperature control in complex volumes of heterogeneous media. The proposed framework provides a solid foundation for anticipating and assessing thermal changes within materials using only outer surface measurement since it can correlate with surface temperature data alone. 

\end{abstract}

\keywords{Acousto-thermal patterning, Sound manipulation, Acoustic holographic lens, Thermal holograms, Acoustic holography} 
\maketitle
Unwanted temperature rise typically presents problems for engineers, whereas controlled heating fills a useful role in applications such as material characterization \cite{jones2014tempature}, defect detection \cite{ibarra2009comparative}, additive manufacturing \cite{buchanan2019metal}, and thermal therapy in medical field \cite{moros2012physics}. Although there are many ways to manipulate heat, focused ultrasound (FUS)-induced thermal fields stand out in particular when it comes to the development of controlled drug-delivery systems \cite{han2013high,needham2001development} and therapeutic effects \cite{jenne2012high} in medical applications, as well as the remote actuation and multi-stimulus control of ultrasound-sensitive smart materials like shape memory polymers (SMPs) \cite{peng2020interaction,bhargava2019coupling, bhargava2017focused}. When exposed to external stimuli such as heat, these polymers can store a temporary shape and revert to their permanent or original shape. FUS has been employed as a noninvasive trigger for stimulating SMP-based systems and offers a higher ability to localize the heating action, allowing the shape recovery process to begin only in specific areas of the polymer. Curved single-element and phased-array transducers are widely used for concentrating ultrasound energy. However, acoustic holographic lenses (AHLs), also known as acoustic holograms, have recently become a hot topic for ultrasonic manipulation in order to construct complex FUS field \cite{melde2016holograms,10.1063/5.0009829,10.1063/1.5048601,AHMED2021}. AHLs, in their most basic form, record the phase profile of the required wavefront, which is then utilized to rebuild the acoustic pressure field when lit by a single acoustic source. Overall, the present popularity of acoustic holograms originates from their ease of use, durability, and low cost, giving them an alternate option for a variety of biomedical operations \cite{9548786, jimenez2019holograms, kim2021acoustic, randad2020design, sallam2023noninvasive,sallam2022nonlinear, sallam2021holographic}. From the thermal field mapping and controlling standpoint, there has been relatively limited research on the use of AHLs to induce specific temperature fields \cite{andres2022thermal,andres2023holographic}. The primary challenge in creating acousto-thermal fields with AHLs arises from the non-homogeneity of the physical domain, which typically consists of a primary medium (e.g., water) and a secondary attenuating medium (e.g. polymers, tissue phantoms, sound absorbing sheets, etc.). To produce the desired pressure field pattern, modeling heterogeneity involves computationally costly three-dimensional full-wave simulations with a number of iterations. Furthermore, the problem's complexity increases with the presence of mixed acousto-thermal effects. The fundamental process of reconstructing diffraction-limited acoustic patterns requires both forward and backward propagation of the pressure field, as well as the imposition of particular boundary conditions. This is in contrast to the direct application of thermal patterns as a temperature rise in full-wave simulations, which results in unstable and non-unique solutions due to the mathematical ill-posedness of inverse heat transfer issues \cite{ozisik2018inverse}.

In this work, we present an effective and precise method for creating AHLs for generating specific and controllable acousto-thermal fields. The volumetric acousto-thermal analysis of thin and thick case studies is carried out to gain a better understanding of the adaptability of AHLs for constructing tailored thermal patterns at depth. Our analysis gives a thorough explanation of the dynamics and patterns seen, as well as demonstrates the utility of holographic FUS thermal mapping for accurate temperature control in complex volumes of heterogeneous media. 

The time-reversal method using three-dimensional k-space pseudospectral linear acoustical simulations is used to generate the required pressure field for constructing the desired target pattern. The selected target pattern is initially imposed as a pressure field at the target plane (\(h_t\)), while the propagating wave is recorded by the sensor layer at the hologram plane's top surface. The collected time-varying pressure field is then reversed in time to extract frequency domain phase (\(\phi_s\)) and amplitude (\(P_s\)) information using a windowed and zero-padded Fast Fourier Transform (FFT).

In the second step, the acquired pressure field is forward propagated from the hologram plane to calculate the pressure amplitude inside the volume of the target medium. The absorbed acoustic field is then converted into heat deposition to determine the temperature rise. The frequency power law, as shown in Eq. \ref{eq:absorption}, governs sound absorption.
\begin{equation}
\alpha=\alpha_0\\f^\gamma\  
\label{eq:absorption}
\end{equation}

\begin{figure*}
    \centering
    \includegraphics[width=1\linewidth]{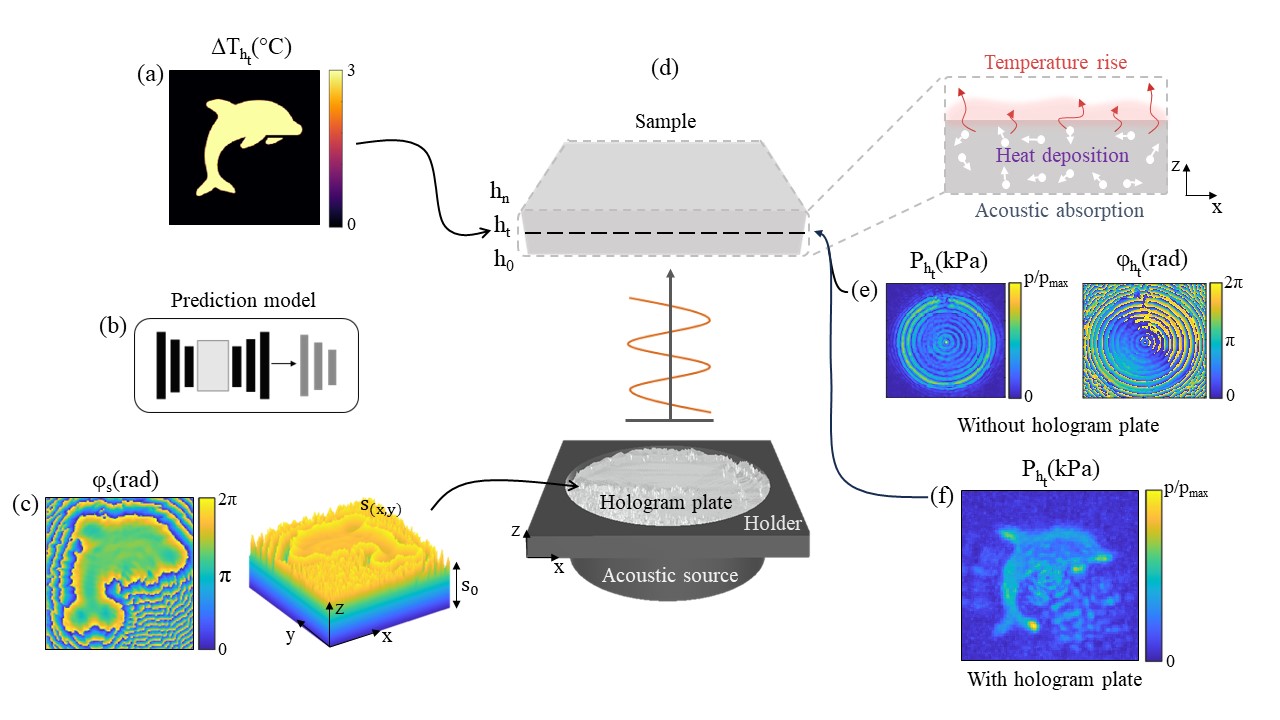}
    \caption{The holographic acousto-thermal patterning workflow. (a) Selected thermal pattern for the target layer at depth (\(h_t\)). (b) Illustration of the deep learning-based model for acoustic source prediction. (c) The output of the source phase pattern (left) and generated thickness map (right). (d) A holographic ultrasound system for acoustic to thermal transformation is depicted schematically. (e) Measured acoustic pressure amplitude (\(P_s\)) (left) and phase (\(\phi_s\)) (right) at the target layer (\(h_t\)) without the hologram plate. (f) Measured acoustic pressure amplitude at the target layer (\(h_t\)) with hologram plate (experimental pressure field measurements are conducted in deionized water only).}
    \label{fig:general_Fig}
\end{figure*}
Here the absorption coefficient (\(\alpha_0\)) is set to \SI{4.2}  
 {dB/MHz} for the commercially available soft silicone rubber: Smooth-On Ecoflex 00-10 \cite{smooth10ecoflex}, widely employed as tissue mimicking material in the literature \cite{adams2017soft, Bowen, estermann2020quantifying}. The absorption coefficient is defined as \SI{0.0022} {dB/MHz} for the surrounding water medium, and the power law absorption exponent (\(\gamma\)) of 1.1 is chosen for the whole domain. Excitation frequency (\(f\)) of 1 MHz is selected for the source according to the typical frequency range for therapeutic ultrasound applications \cite{speed2001therapeutic}. Additionally, for water and silicone rubber, the density and speed of sound are adjusted as follows \(\rho_w\)=\SI{1000}{kg/m^3}, \(c_w\) = \SI{1481}{m/s}, \(\rho_t\)= \SI{1040}{kg/m^3} \cite{smooth10ecoflex} and \(c_t\) = \SI{1020}{m/s} \cite{mark2007physical}, respectively. The computational domain consists of 350x350 pixels, with a grid size of 0.2 mm. Additionally, the domain is expanded with perfectly matched layers (PMLs) to prevent undesired reflections from model boundaries. 

The volumetric acousto-thermal analysis of the sample material is conducted layer by layer at depth throughout the entire paper. The initial layer (\(h_0\)) depicts the separation between the sample and the hologram plate's top surface. The uppermost surface of the sample is indicated by (\(h_n\)), and the target layer on which a particular pattern is imposed is designated by (\(h_t\)). The overall sample material thickness is determined by the difference between the top and bottom layers (\(h_n-h_0\)), Fig.\ref{fig:general_Fig}(d).

The volumetric heat deposition (\textit{q}) in the sample material as a result of pressure field attenuation  (\textit{p}) owing to absorption is calculated using the relation \cite{bailey2003physical}:
\begin{equation}
q=\alpha\\p^2/\rho_t\\c_t 
\label{eq:heat deposition}
\end{equation}

Following that, the resulting temperature rise is then calculated using Penne's Bio Heat equation \cite{pennes1948analysis}:

\begin{equation}
\rho_t C_t \frac{\partial T}{\partial t} = \nabla \cdot (\kappa \nabla T) + q_b + q \label{eq:bioheat}
\end{equation}
Because this study uses a soft silicone rubber with no fluid flow inside, convection due to blood flow (\(q_b\)) is neglected, and only heat deposition and diffusion effects are taken into account for estimating temperature fluctuations. Thermophysical properties of the sample material are set to \(\kappa=\) \SI{0.16}{W/mK} \cite{bhanushali2017copper} and \(C_t=\)\SI{1558}{J/kgK} \cite{forspecificheat} in Eq.\ref{eq:bioheat} for calculating thermal conductivity and specific heat capacity, respectively.

To generate precisely targeted acousto-thermal patterns, discrepancies between the imposed and forward-propagated pressure fields necessitate iterative simulations and processing efforts. In a variety of applications, deep learning techniques have been offered as a solution to inverse heat and design problems \cite{tamaddon2020data, zhu2022deep, garcia2021deep, xi2023ultrahigh, azarifar2022machine, hu2020machine}. We use a machine learning system to compress the process into a single inverse issue for speedy and efficient holographic plate design in this study. To assess the relevant input data and generate the necessary output data, a data-driven deep learning (DL) network with U-Net architecture composed of encoder and decoder blocks is used. This method entails converting thermal information into a holographic representation by utilizing two latent functions: pressure phase and amplitude. The training set for the network was generated by an in-house written MATLAB code, employing the aforementioned acousto-thermal modeling approach. For adversarial training, the discriminator model uses binary cross-entropy (BCE) loss calculations to quantify differences between the ground truth and synthesized pressure maps:

\begin{equation}
BCE=-(GT)log(\hat{GT})+(1-GT)log(1-GT)
\end{equation}

Here \textit{GT} represents the ground truth and \(\hat{GT}\) the prediction probability. Aside from BCE, the mean absolute error (MAE) is used for pixel-wise error assessment to evaluate the generator model's performance for predicting (\textit{Est}) the desired pressure field information. Once a workable DL algorithm has been created, the desired pressure field information can be retrieved by feeding the model the selected thermal pattern. 

\begin{equation}
MAE=\frac{1}{n_{step}} \sum_{m=1}^{n_{step}} \left| GT_m - Est_m \right|
\end{equation}

The next step is to transform the DL model's output \(\phi_s\)  and \(P_s\) into a Phase-Only-Hologram (POH) to build the final thickness map for 3D printing. This is accomplished through the use of bidirectional error diffusion (BERD) \cite{tsang2013novel} for phase modification. In contrast to the original approach, which used a constant amplitude of 1 for the raw pixels, we used a pattern-adaptive optimization parameter to improve the quality of the empirically reconstructed image \cite{liu2021pattern}. Finally, the thickness map for 3D printing is calculated using the following formula \cite{melde2016holograms}:

\begin{equation}
s_{(x,y)}=s_0-\frac{\Delta\phi_{(x,y)}}{k_w-k_h}
\end{equation}

Here, each pixel's thickness in the hologram plane is indicated by  \(s_{(x,y)}\), the phase-lag between source and POH is represented by \(\Delta\phi_{(x,y)}\), and the hologram plate's initial thickness is specified as \(s_0\). The wave numbers for hologram material and surrounding medium are defined as \(k_h\) and \(k_w\), respectively.

\begin{figure}
    \centering
    \includegraphics[width=1\linewidth]{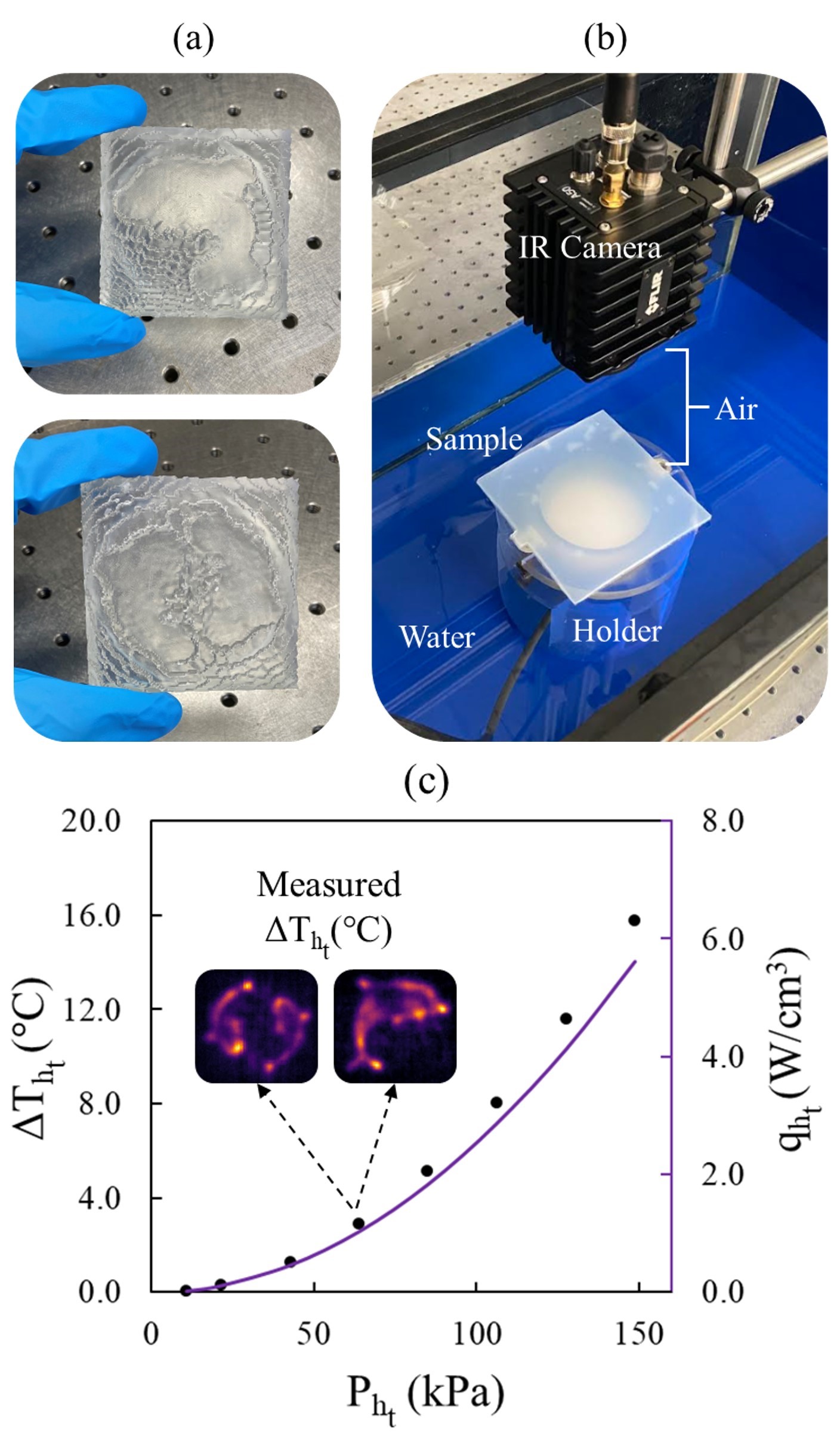}
    \caption{(a) 3D printed hologram plates for testing. (b) Experimental setup for the thermal measurements. (c) Relationship at the target layer for the thin sample with a 1-mm thickness between heat deposition, temperature rise, and acoustic pressure amplitude. Additionally, a reference measurement point for the experiment is shown.}
    \label{fig:exp}
\end{figure}

AHLs are fabricated through stereolithography (SLA) techniques using the Form 2 resin printer by Formlabs; capable of achieving a printing resolution ranging from \SI{25}{\micro\meter}-\SI{100}{\micro\meter}. Acoustic properties of the photo-polymer (Clear Resin) are \(\rho_h =\) \SI{1178}{kg/m^3}, \(c_h=\) \SI{2594}{m/s} and \(\alpha_0=\) \SI{2.92}{dB.MHz^{-y} .cm^{-1}} \cite{bakaric2021measurement} and the final thickness of the lens is 6 mm, Fig. \ref{fig:exp}(a). Detailed discussions on the calculated transmission power coefficient and acoustic pressure field efficiency of the system, both with and without AHL, are provided in Note S1-S2. The resulting transmission coefficient distribution and efficiency maps can be found in Figs. S1-S3 of the supplementary Material.

Ecoflex 00-10 testing samples are made by combining two liquid ingredients in a 1:1 ratio. The liquid mixture is injected into a controlled-height silicone mold to achieve the appropriate sample thickness (\(h_n-h_0\)) and cured at room temperature. For the experimental testing, two case studies are studied: \(h_n-h_0=\)\SI{1}{mm} (thin) and \(h_n-h_0=\)\SI{30}{mm} (thick) samples. 

\begin{figure*}
     \centering
    \includegraphics[width=1\linewidth]{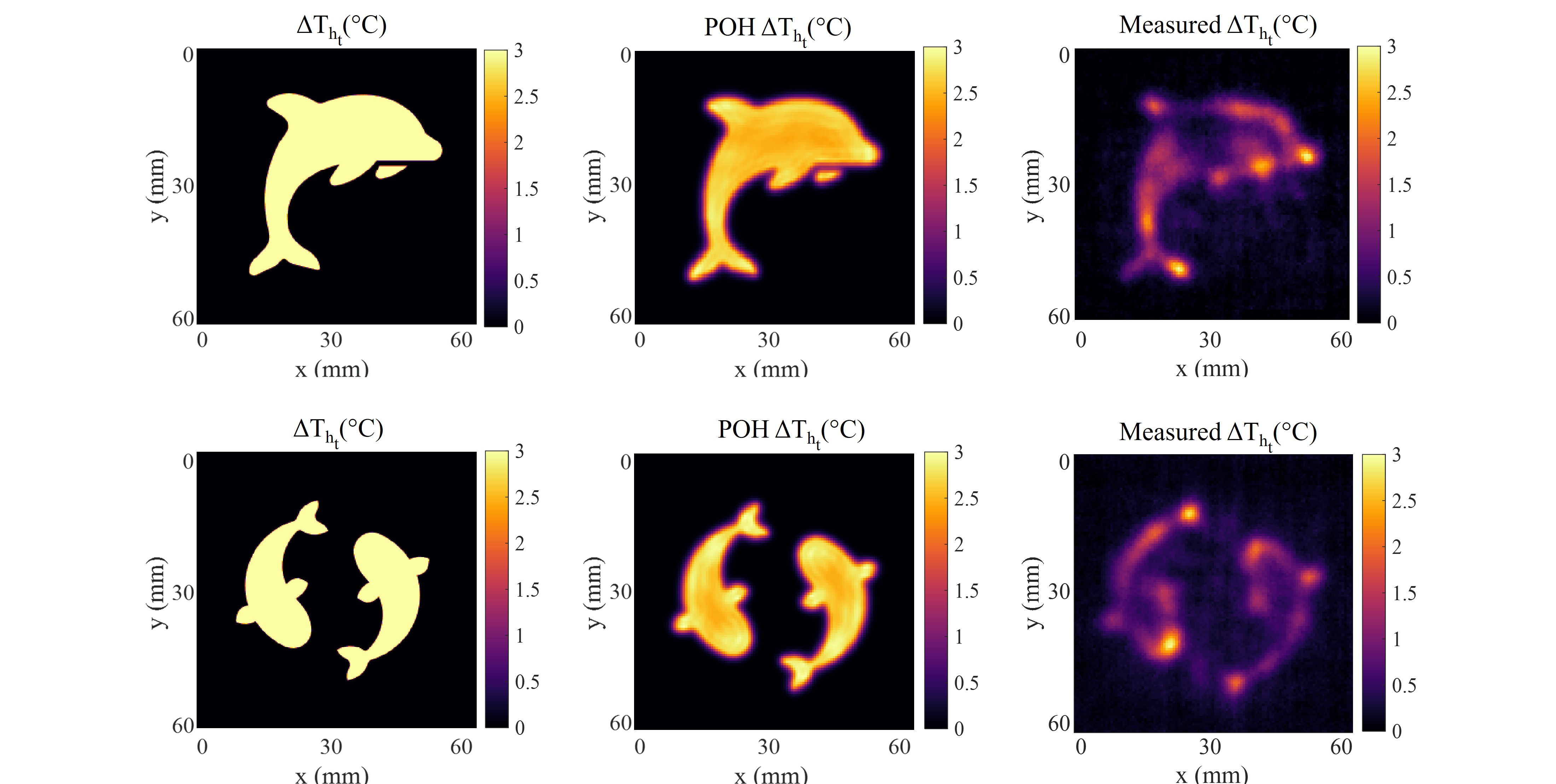}
    \caption{Thermal maps for two different patterns with a continuous ultrasonic exposure period of 10 seconds. \(\Delta T_{h_t}\), POH \(\Delta T_{h_t}\) and Measured \(\Delta T_{h_t}\) represent input thermal maps to the DL model, simulation and experimental results respectively.}
    \label{fig:thermalres1} 
\end{figure*}

The study's experimental verification is divided into two steps: pressure and thermal measurements. Both experiments are carried out in a tank of degassed and deionized water. Precision Acoustics, Ltd.'s Aptflex F28 acoustic absorbing sheets are used to ensure acoustic isolation in a section of the tank. The transducer and acoustic hologram are held in place by an in-house holder. The acoustic source is a flat transducer made of a 50 mm diameter piezoelectric disk that operates at 1 MHz. The acoustic source is excited with a sinusoidal burst signal consisting of 20 cycles at 10 ms intervals for pressure field measurements. A signal generator (Keysight 33500B) and an amplifier (E\&I A075) power the transducer. Pressure amplitude mapping is performed in the initial studies using a needle hydrophone (ONDA HNR-0500) connected to a 3D positioning system. A calibration factor of  -252.81 dB (re. \SI{1}{V/\micro P} at 1 MHz), provided by the manufacturer, is used to convert the voltage data into acoustic pressure information. An oscilloscope (Tektronix TBS2104) is used to visualize the sampled digital signal, and the recorded signals are post-processed using a MATLAB software. To cover the complete pattern at the target plane, a 60x60 mm scanning zone with a 0.5 mm step size is chosen. 
(\(h_t =\) 23 mm). The resulting pressure field measurements are given in Figs. \ref{fig:general_Fig}(e) and (f). A clear dolphin pattern conversion is clearly visible when the transducer pressure field, Fig. \ref{fig:general_Fig}(e), is compared with the hologram incorporated pressure amplitude distribution, Fig.\ref{fig:general_Fig}(f), at the target plane. While a satisfactory normalized pressure field pattern is obtained, undesirable pressure amplitudes are found in the outer areas. The hologram plate was designed to create a thermal pattern on a sample material, whereas hydrophone scans were only performed in the water domain. Overall, hydrophone measurements provide useful insights about the holographic plate's performance prior to thermal measurements, allowing us to adjust the transducer's operational settings for acousto-thermal conversion, Fig.\ref{fig:exp}(c). 

An upgraded mobile part is added to the holder design in the second phase of the experimental arrangement, Fig.\ref{fig:exp}(b), to secure the sample material at a certain distance \(h_0\) from the hologram surface. Infrared (IR) thermal camera (FLIR A50-51°, Teledyne FLIR LLC) with IR resolution of 464 × 348 and thermal sensitivity of \(<\) \SI{35}{mK} \cite{estermann2020quantifying} is employed to measure the thermal distribution. Following transducer activation, the recording length is set to 10 seconds, with 30 frames taken each second. Thermal pictures are also preserved for later post-processing, beginning with thermal equilibrium. Furthermore, the region between the IR camera and the upper surface of the sample is retained as air to prevent any undesired IR emission by water. The measurements were taken at room temperature and humidity. The emissivity of the silicone rubber is set as \( \epsilon=0.9\) for the measurement.

Thermal measurement results are compared in Fig. \ref{fig:thermalres1} for two different selected patterns (dolphin and yin-yang fish). The first experimental analysis is done for the \(h_n-h_0 =\) \SI {1} {mm} thin sample where \(h_n=h_t=\) \SI{23}{mm}. The results show that a maximum temperature difference of 3°C is achieved across all cases for the 10-second heating time and the proposed machine learning-assisted hologram production approach successfully forms specified target patterns at \(h_t\). 
 
\begin{table*}
\renewcommand{\arraystretch}{2}
\caption{Error calculation results for the simulated and measured thermal maps with respect to the imposed thermal field. High PSNR, low RMSE and a high SSIM is indicative of high quality.}
\label{tab:error}
\begin{ruledtabular}
\begin{tabular}{ccccc}

{} & {} & PSNR (dB) $\uparrow$ & RMSE $\downarrow$ & SSIM $\uparrow$ (1-0) \\ 
 \hline 
\multirow{-2}{*} & POH $\Delta T_{h_t}$ (°C) & 13.95 & 0.20 & 0.76 \\
\multirow{-2}{*}{\includegraphics[scale=0.4, valign=c]{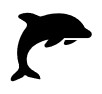}} 
 & Measured $\Delta T_{h_t}$ (°C)& 12.73 & 0.23 & 0.33 \\
 
\multirow{-2}{*}&POH $\Delta T_{h_t}$ (°C) & 13.95 & 0.20 & 0.76 \\
\multirow{-2}{*}{\includegraphics[scale=0.4, valign=c]{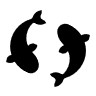}} 
& Measured $\Delta T_{h_t}$ (°C) & 10.73 & 0.29 & 0.32 \\ [1ex] 

\end{tabular}
\end{ruledtabular}
\end{table*}

 \begin{figure*}
    \centering
    \includegraphics[width=1\linewidth]{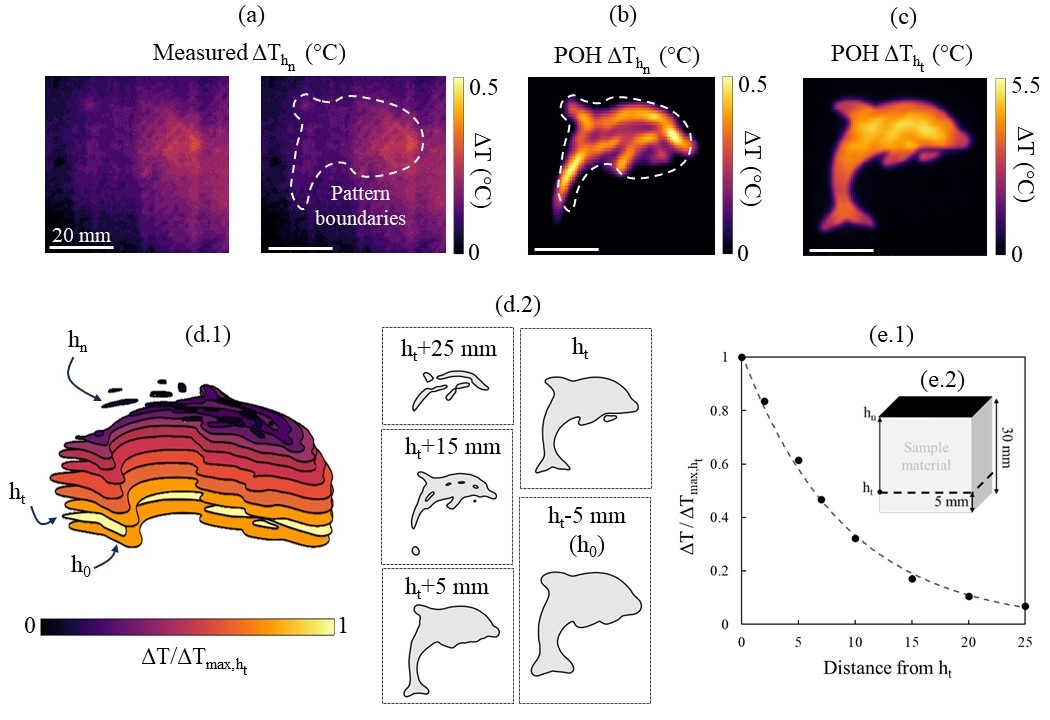}
    \caption{Volumetric thermal pattern for the 30-mm thick sample. (a) IR camera measurement for the top surface (\(h_n\)). For visual comparison, pattern boundaries are indicated. The simulation thermal map for layers (b) \(h_n\) and (c) \(h_t\).
(d.1) Volumetric representation of the layer-by-layer pattern variation and normalized temperature rise \(\Delta T\)/\(\Delta T_{max,h_{t}}\). (d.2) Weighted average temperature rise is converted into a binary map for pattern comparison. (e.1) Temperature variation in simulation data normalized by distance from the target layer. (e.2) The layer distances inside the sample material are depicted schematically.}
    \label{fig:thick all}
\end{figure*}

A comparison of error metrics is presented in Table \ref{tab:error}, with the corresponding equations available in the supplementary material Note S3. The rooth mean squared error (RMSE) values for both the simulation and experimental thermal maps are low, indicating a tight match between the imposed and acquired temperature elevation. The average peak signal-to-noise (PSNR) value across all four examples is 12.84 dB, with the simulated POH \(\Delta T\) obtaining somewhat higher precision for thermal patterning. Similarly, the simulated thermal maps had significant SSIM values, indicating a remarkable pattern similarity to the imposed thermal map.  Furthermore, the average structural similarity index (SSIM) of 0.325 for the measured \(\Delta T\) indicates that clear patterns can be achieved experimentally.
Given the experimental errors, such as 3D printing tolerances for the hologram plate, material imperfections, and unwanted IR reflections during thermal camera measurements, machine learning-generated holograms offered a promising solution to fast and accurate acousto-thermal patterning.

We investigate acousto-thermal patterning within a thicker sample material in addition to the single-layer study, \(h_n-h_0=\) \SI{30}{mm}, Fig. \ref{fig:thick all}. Following a similar technique to the single-layer patterning methodology, a new hologram plate is carefully designed to target a specific layer within this thicker sample. Figure \ref{fig:thick all}(d.1)-(d.2) provides a detailed layer-by-layer examination of the thermal pattern variation at depth, while the graph in Fig. \ref{fig:thick all}(e.1) illustrates the normalized maximum temperature variation as the distance from the target layer changes. Weighted average temperature variation is obtained for each layer and translated into a binary map for a more accurate comparison of pattern behavior, as illustrated in Fig. \ref{fig:thick all}(d.2). Temperature rise is normalized to the target layer data's local maximum and assigned to the binary map in Fig. \ref{fig:thick all}(d.1), for three-dimensional representation.

The results indicate a well-defined dolphin pattern at the chosen target layer, with a noticeable increase in heat buildup towards the pattern's center as one goes away from the target surface, coupled by a drop in the maximum temperature rise. Heat deposition owing to ultrasonic absorption rapidly decreases about \(h_t\)+\SI{15}{mm} and above, resulting in a dominance of heat diffusion by thermal conduction inside the volume of the material. Patterns depicted in Fig. \ref{fig:thick all}(d.2) clearly demonstrate this phenomenon, with weighted average temperature rise only evident at the center spots as the distance from the target layer increases, while the rest of the dolphin pattern is not visible due to thermal uniformity. Experimentally, the same behavior is seen, as illustrated in the comparison between  Figs. \ref{fig:thick all}(a)-(b)-(c). (Refer to Note.S4 in the supplementary material for additional surface measurements across various thicknesses.) When studying the pattern boundaries, the top layer thermal map comparison shows a close match between simulation and experiment data. For 10 seconds of ultrasound exposure time, a temperature rise up to 0.5°C is observed at the \(h_n\). As indicated in the graph in Fig.\ref{fig:thick all}(c), this nearly corresponds to a 5.5°C temperature rise at the \(h_t\) where a well-defined dolphin pattern is visible. 

Our method's ability to correlate with surface temperature data alone, for the first time, offers a strong foundation for predicting and analyzing AHL induced thermal changes within materials using only outer surface measurement. Further investigation into this feature of our methodology has the potential to generate practical benefits and could be a future research topic of interest. 

A novel machine learning-assisted holographic approach for acousto-thermal patterning in soft material samples was proposed in this study. We presented a thorough analysis that includes both numerical calculations and experimental validations to demonstrate how 3D-printed AHLs may be utilized to modulate sound to create heat maps. We compared two different arbitrary target shapes as well as major image quality parameters. During the final step of our research, we focused on volumetric analysis and layer-by-layer pattern variation inside a thick soft sample. We were able to establish a clear thermal pattern at the target layer, and as we moved away from the target zone, the thermal effect and pattern clarity gradually faded. Pattern focusing was achieved by combining a simple flat transducer with an acoustic hologram without the need for a curved or phased array transducer. The proposed acousto-thermal modeling enables applying precise volumetric temperature control of sample materials which is crucial for studying thermal phenomena and optimizing for broad application. Moreover, with the AHL's easy adaptability introduced through the acousto-thermal based approach, holographic lens design can be expanded to address multiple focal points and volumes simultaneously. Notably, the framework, which can correlate with surface temperature data alone, offers a solid platform for forecasting and evaluating thermal changes within materials using only outer surface measurement for the first time. Because of their specific features and capabilities, AHLs hold considerable promise for developing volumetric thermal holography for future applications in therapeutic ultrasound, material characterization, additive manufacturing, and noninvasive actuation of ultrasound-sensitive smart materials.

\hspace{1cm} 

See supplementary material for acoustic power transmission calculations (Note S1 and Eq.S1); corresponding distribution for the AHL (Fig.S1); for details of the pressure field efficiency analysis (Note S2 and Eqs.S2-S3); corresponding pressure magnitude and efficiency reduction (Figs.S2-S3); Equations for error metric calculations (Note S3); Details of IR Camera measurements for volumetric validation (Note S4); and corresponding thermal maps (Fig.S4).

\hspace{1cm} 

The data that support the findings of this study are available from the corresponding author upon reasonable request.

\begin{acknowledgments}
This work was supported by the U.S. National Science
Foundation (NSF) under the grants, Award No.
CAREER CMMI 2143788, and CMMI 2016474,  which are gratefully acknowledged.
\end{acknowledgments}

\bibliography{ref}% Produces the bibliography via BibTeX.
\end{document}

\title{Supplementary Material\\Holographic Thermal Mapping Using Acoustic Lenses}

\author{Ceren Cengiz}
%\email{cerencengiz@vt.edu}
\author{Shima Shahab}
\email{sshahab@vt.edu}
\affiliation{Department of Mechanical Engineering, Virginia Tech, Blacksburg, VA, $24061$, USA}
\maketitle

\section*{Note S1. Acoustic power transmission calculations}
Sound propagation through the acoustic holographic lens (AHL) involves multiple mediums, leading to variations in transmitted acoustic power compared to the input power due to impedance differences. To quantify the impact of impedance mismatch, a transmission coefficient is computed using Eq.S1. In Eq.S1, \(Z_{t}=c_{t}p_{t}\), \(Z_{h}=c_{h}p_{h}\) and \(Z_{m}=c_{m}p_{m}\) represent the acoustic impedance of the transducer surface, hologram, and propagating medium, respectively. Here, s(x,y) corresponds to the generated thickness map derived from phase differences, and \(k_{h}=2\pi f/c_{h}\) denotes the wavenumber for the holographic lens. The calculated transmission coefficient is provided as a reference in Fig.S1.
\setcounter{equation}{0} % Reset figure counter
\renewcommand{\theequation}{S\arabic{equation}} % Prefix figure 
\begin{equation}
    TR(x,y)=\frac{4Z_{t}Z^2_{h}Z_{m}}{Z^2_{h}(Z_{t}+Z_{m})^2cos(k_{h}s(x,y))^2+(Z^2_{h}+Z_{t}Z_{m})^2sin(k_{h}s(x,y))^2}
\end{equation}

\setcounter{figure}{0} % Reset figure counter
\renewcommand{\thefigure}{S\arabic{figure}} % Prefix figure numbers with "S"
\begin{figure}
    \centering
    \includegraphics[width=0.6\linewidth]{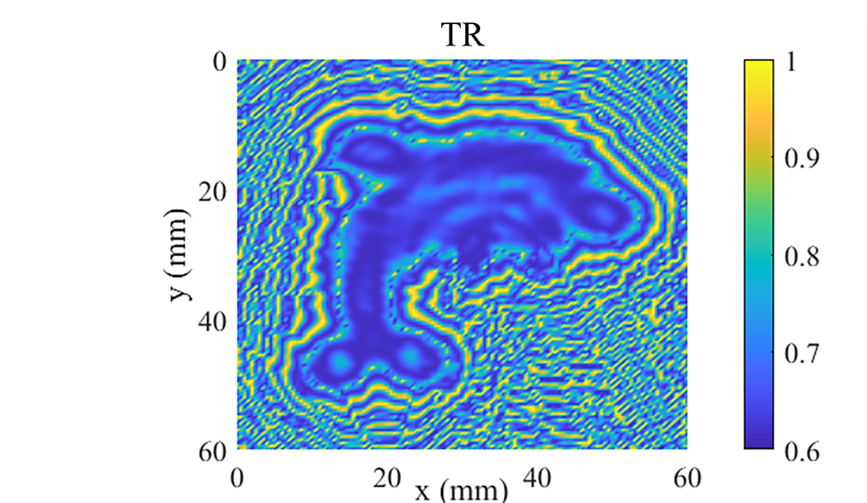}
    \caption{Corresponding acoustic power transmission coefficient distribution for the dolphin pattern.}
    \label{fig1}
\end{figure}

\section*{Note S2. Experimental analysis of pressure field variation in the presence of AHL}

The section focuses on experimental analysis, specifically investigating the variation in the pressure field with the inclusion of AHLs. A custom-built transducer with a diameter of 50 mm is actuated by a sinusoidal burst signal composed of 20 cycles at 10 ms intervals, employing a Keysight 33500B signal generator and an A075 E\&I power amplifier. Pressure measurements are undertaken using an ONDA HNR-0500 hydrophone. Figure S2 illustrates the outcomes of pressure field measurements. The hydrophone is positioned 29 mm from the transducer surface, with the target layer situated at a distance of 23 mm from the hologram plane, and the lens thickness being 6 mm. 

\begin{figure}[H]
    \centering
    \includegraphics[width=0.8\linewidth]{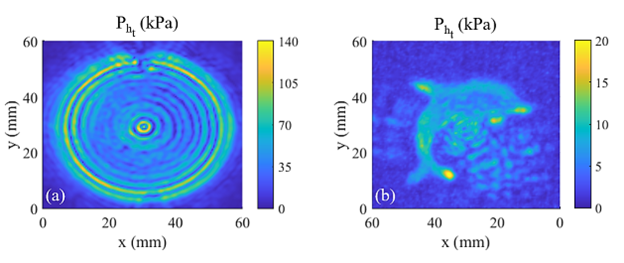}
    \caption{Hydrophone experimental pressure field measurements at the target plane (a) without the hologram (b) with the hologram. Note: The quality of the observed pattern may not accurately represent the actual pattern, as it is designed for use within a tissue-mimicking material rather than in water alone. For a comprehensive analysis of the pattern performance, refer to Figs.3-4 and Table.1 in the manuscript.}
    \label{fig:enter-label}
\end{figure}

Figure S2 illustrates pressure field variations with the addition of the hologram. In Fig.S2(a), without the hologram, the transducer produces a diffused distribution with a 140 kPa magnitude at the target plane. Whereas Fig.S2(b) depicts a 120 kPa pressure decrease at the target plane when the hologram is introduced.

To quantify the significance of the pressure field drop introduced by the AHL, a custom Figure of Merit (FOM) is defined in Eq.S2. Given that optimizing efficiency requires prioritizing reductions pertinent to the desired pattern, any decreases occurring outside this pattern are deemed irrelevant to performance. Consequently, a weighted FOM value, as defined in Eq.S3, is introduced to selectively account for areas within the specified target region.

\begin{equation}
FOM(i,j)=\frac{(P_{AHL}(i,j)-P_{source}(i,j))}{(\max(P_{source}))} \times 100
\end{equation}
\begin{equation}
FOM_{wt}(i,j)=\frac{w(i,j)(P_{AHL}(i,j)-P_{source}(i,j))}{(\max(P_{source}))} \times 100
\end{equation}
\hspace{1cm} 

\(P_{AHL}(i,j)\) and \(P_{source}(i,j)\) represent the pressure field elements in the matrix with and without the hologram, respectively (refer to Fig. S2.). In Eq.S3, w(i,j) denotes the weight assigned to exclude regions outside the matrix elements at (i,j) of the pattern. The resulting FOM distributions are shown in Fig. S3. 

\begin{figure}[H]
    \centering
    \includegraphics[width=0.8\linewidth]{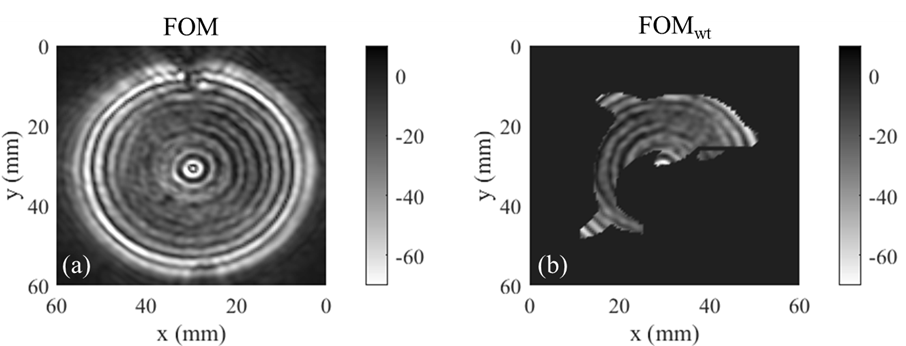}
    \caption{Efficiency evaluation through FOM calculations: (a) Overall FOM for the entire scanning region, and (b) weighted FOM focusing specifically on the pattern region.}
    \label{fig:enter-label}
\end{figure}
As expected, an efficiency reduction is observed when AHL is added to the system (Fig.S3(a)). However, Fig.S3(b) clearly highlights the fact that for the area of interest, this decrease is much lower, with some areas showing improved efficiency. Overall, the average pressure efficiency decrease in the target region is calculated to be 21\% in Fig.S3(b). These correspond to a maximum reduction of 10\%  and an average of 1.1\%  in the temperature field according to the acoustic absorption and heat diffusion calculations given in the paper. 

\section*{Note S3. Error calculations for comparison of experimental and simulation thermal maps}
For the error metric calculations, the following equations are used:

\begin{equation}
    PSNR=10log10\frac{MAX^2}{MSE}
\end{equation}
\begin{equation}
MSE = \frac{1}{ab} \sum_{i=1}^{a} \sum_{j=1}^{b} \left[ (\hat{\Delta T}(i,j) - \Delta T(i,j) \right]^2
\end{equation}
\begin{equation}
    RMSE=\sqrt{MSE}
\end{equation}
\begin{equation}
    SSIM = \frac{(2\mu_x \mu_y + L_1)(2\sigma_{xy} + L_2)}{(\mu_x^2 + \mu_y^2 + L_1)(\sigma_x^2 + \sigma_y^2 + L_2)}
\end{equation}

Here MAX is the maximum pixel value, which can be evaluated using \(2^D-1\), where \(D\) is the dynamic range of either the measured \(\Delta T\) or POH \(\Delta T\). In Eq. S5, \(\Delta T\) represents the ground truth thermal map, while \(\Delta T\) is the numerical or the experimental thermal map results. Rows and columns of the data matrix are denoted by \(i\), \(a\) and \(j\), \(b\) respectively. In the calculation of \(SSIM\) (Eq. S7), \(\mu_x\) and \(\mu_y\) denote the local mean values, \(\sigma_x^2\) and \(\sigma_y^2\) represent the variances, and \(L_1\) and \(L_2\) are the stability constants.

\section*{Note S4. Additional thermal measurements using an infrared (IR) camera to validate volumetric temperature distribution.}

In this section, supplementary infrared (IR) camera temperature measurements were conducted to examine pattern variation in the 30 mm thick sample. Three additional samples with thicknesses of 1 mm, 5 mm, and 15 mm were prepared to assess the formation of the dolphin pattern. It is crucial to note that the tested AHL is specifically designed for the 30 mm thick sample, where the intended target layer (\(h_t\)) is located 5 mm away from the bottom surface \(h_0\). Achieving comprehensive volumetric 3D thermal mapping without disrupting the acousto-thermal field within the sample is challenging. These additional experiments provide valuable insights into the success of AHLs in achieving complex thermal patterns. While these measurements do not directly compare to the volumetric analysis presented in the main manuscript, they serve as an essential supplementary data for validation purposes. As depicted in Fig.S4, the highest-quality thermal pattern is attained when \(h_t=h_n-h_0=5 mm\) resulting in the maximum temperature increase \(\Delta T\). As the difference between \(h_n\) and \(h_t\), the pattern behavior shifts towards thermal uniformity with reduced quality. 

\begin{figure}[H]
    \centering
    \includegraphics[width=0.8\linewidth]{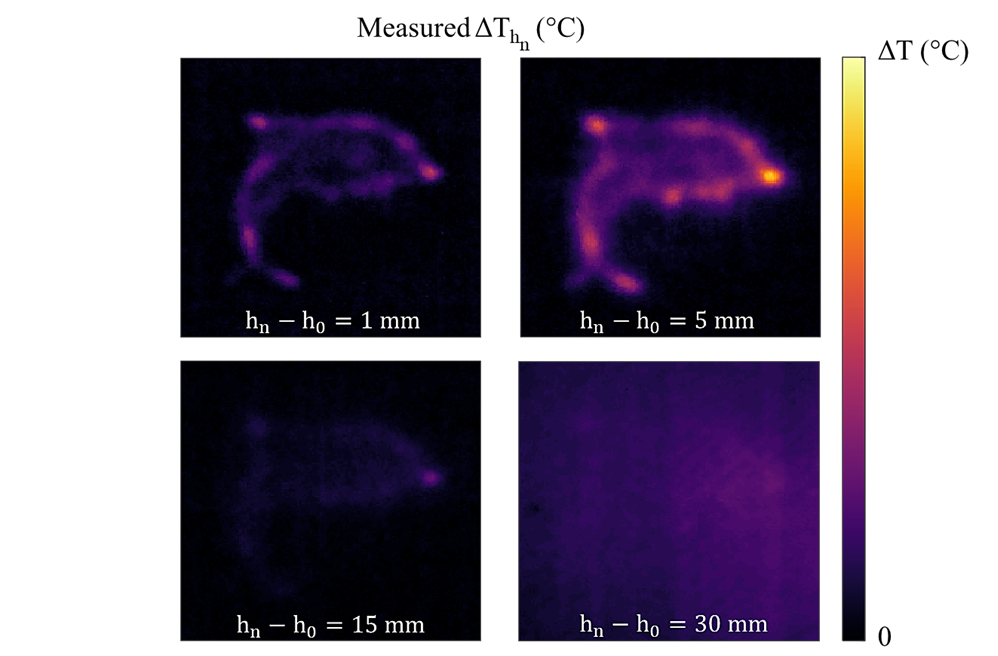}
    \caption{Upper surface Infrared (IR) camera thermal measurements of sample with various thicknesses. Target layer \(h_t\) is at 5 mm while the actual thick sample case is the \(h_n-h_0=30 mm\), where \(h_n\) is the top surface of the sample while \(h_0\) is the bottom surface}
    \label{fig:enter-label}
\end{figure}